\documentclass[plb,12pt]{elsarticle}
\usepackage{graphicx}
\usepackage{dcolumn}
\usepackage{amssymb}

\newcommand{\beq}{\begin{equation}}
\newcommand{\eeq}{\end{equation}}
\newcommand{\beqa}{\begin{eqnarray}}
\newcommand{\eeqa}{\end{eqnarray}}

\journal
{Physics Letters B}

\begin{document}

\begin{frontmatter}
\title{`Excess' of primary cosmic ray electrons}
\author{Xiang Li$^{(a)}$$^{(b)}$, Zhao-Qiang Shen$^{(a)}$$^{(b)}$, Bo-Qiang Lu$^{(a)}$$^{(c)}$, Tie-Kuang Dong$^{(a)}$, Yi-Zhong Fan$^{(a)}$$^{(d)}$\footnote{Corresponding author (email: yzfan@pmo.ac.cn)}, Lei Feng$^{(a)}$\footnote{Corresponding author (email: leifeng@pmo.ac.cn)}, Si-Ming Liu$^{(a)}$, and Jin Chang$^{(a)}$$^{d}$}

\address[a] {Key Laboratory of Dark Matter and Space Astronomy, Purple Mountain Observatory, Chinese Academy of Sciences, Nanjing
210008, China}
\address[b]{University of Chinese Academy of Sciences, Beijing, 100012, China}
\address[c]{School of Physics, Nanjing University, Nanjing, 210092, China}
\address[d]{Collaborative Innovation Center of Modern Astronomy and Space Exploration, Nanjing University, Nanjing, 210046, China
\\}
\begin{abstract}
With the accurate cosmic ray (CR) electron and positron spectra (denoted as $\Phi_{\rm e^{-}}$ and $\Phi_{\rm e^{+}}$, respectively) measured by AMS-02 collaboration, the difference between the electron and positron fluxes (i.e., $\Delta \Phi=\Phi_{\rm e^{-}}-\Phi_{\rm e^{+}}$), dominated by the propagated primary electrons, can be reliably inferred. In the standard model, the spectrum of propagated primary CR electrons at energies $\geq 30$ GeV softens with the increase of energy. The absence of any evidence for such a  continuous spectral softening in $\Delta \Phi$ strongly suggests a significant `excess' of primary CR electrons and at energies of $100-400$ GeV the identified  excess component has a flux comparable to that of the observed positron excess. Middle-age but `nearby' supernova remnants (e.g., Monogem and Geminga) are favored sources for such an excess.
\end{abstract}
\end{frontmatter}

Thanks to the rapid progresses made in measuring the spectra of cosmic ray (CR) electrons and positrons, the presence of significant excesses in both the positron spectrum and the electron/positron total spectrum, with respect to the prediction of standard CR model \cite{Strong07Rev}, has been well established \cite{atic,pamela-positron,AMS-2013,AMS-2014a,AMS-2014b}.
These excesses, attracting great attention, have been widely interpreted as a signal of dark matter annihilation/decay or alternatively the presence of new CR electron/positron sources \cite{fan}. In view of the spectral hardening displayed in the proton and heavier CR particle data of ATIC \cite{lab1}, CREAM \cite{lab2} and PAMELA \cite{lab3}, it is quite natural to speculate that the primary CR electron spectrum also gets hardened at high energies (i.e., there is also an electron excess component, which just accounts for part of the total spectrum excess) and interesting observational signal is expected in AMS-02 data \cite{Feng2014}. The joint fit of the positron-to-electron ratio (${\cal R}=\Phi_{\rm e^{+}}/(\Phi_{\rm e^{+}}+\Phi_{\rm e^{-}})$, where $\Phi$ is the flux) data and the positron/electron total flux data ($\Phi_{\rm tot}=\Phi_{\rm e^{+}}+\Phi_{\rm e^{-}}$) does favor such a possibility \cite{Feng2014,pulsar2013,LinSJ2015}. However, in the model of multiple pulsars for the positron excess \cite{Yin2013} the primary-electron spectrum hardening/excess is found to be not needed. Such a ``divergency" demonstrates that  it is necessary to ``identify" the excess as model-independent as possible, which is the main goal of this work.

\begin{figure}
\includegraphics[width=75mm,angle=0]{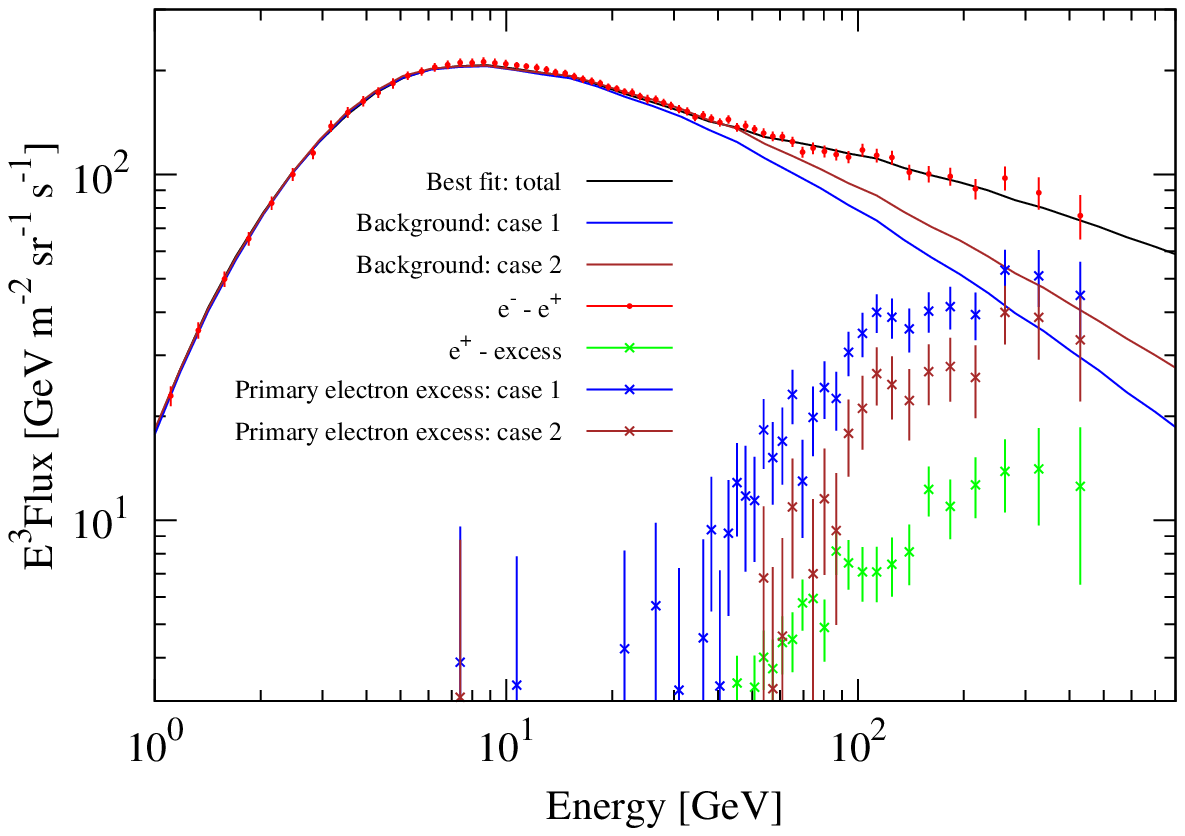}
\includegraphics[width=75mm,angle=0]{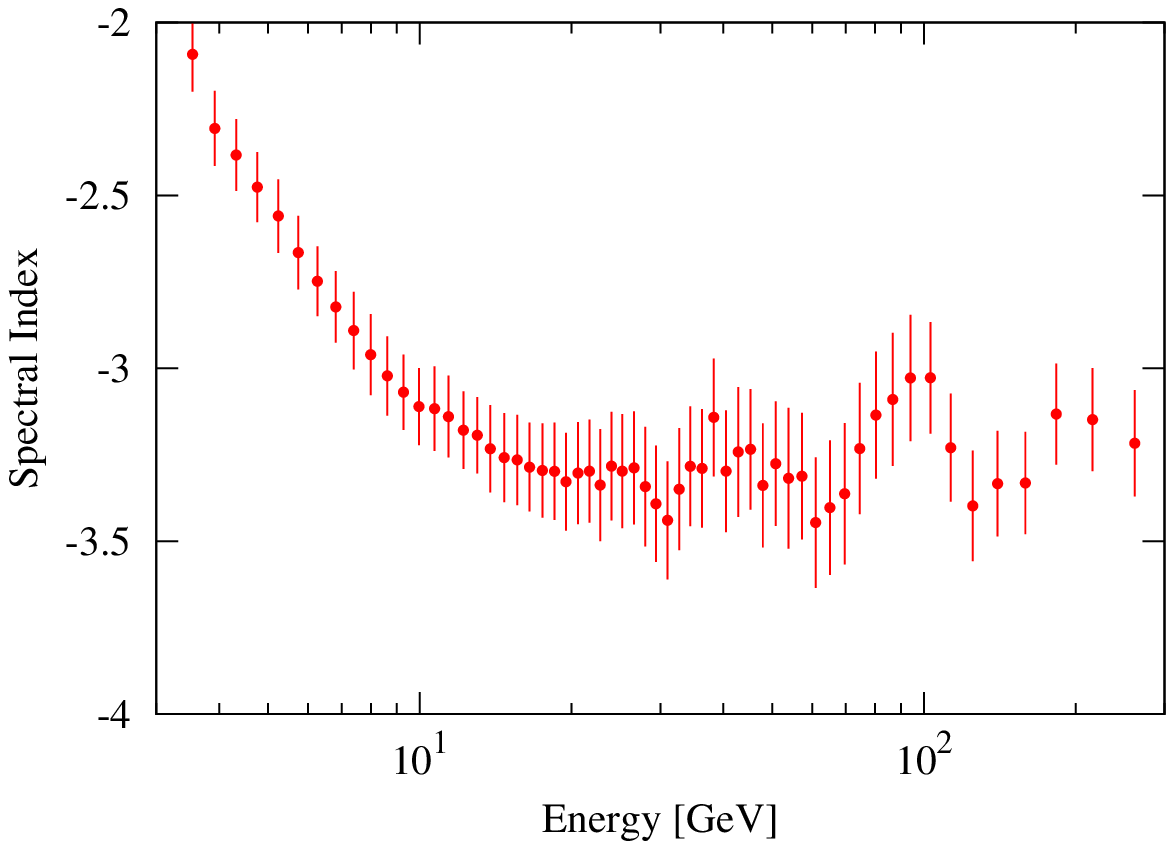}
\includegraphics[width=75mm,angle=0]{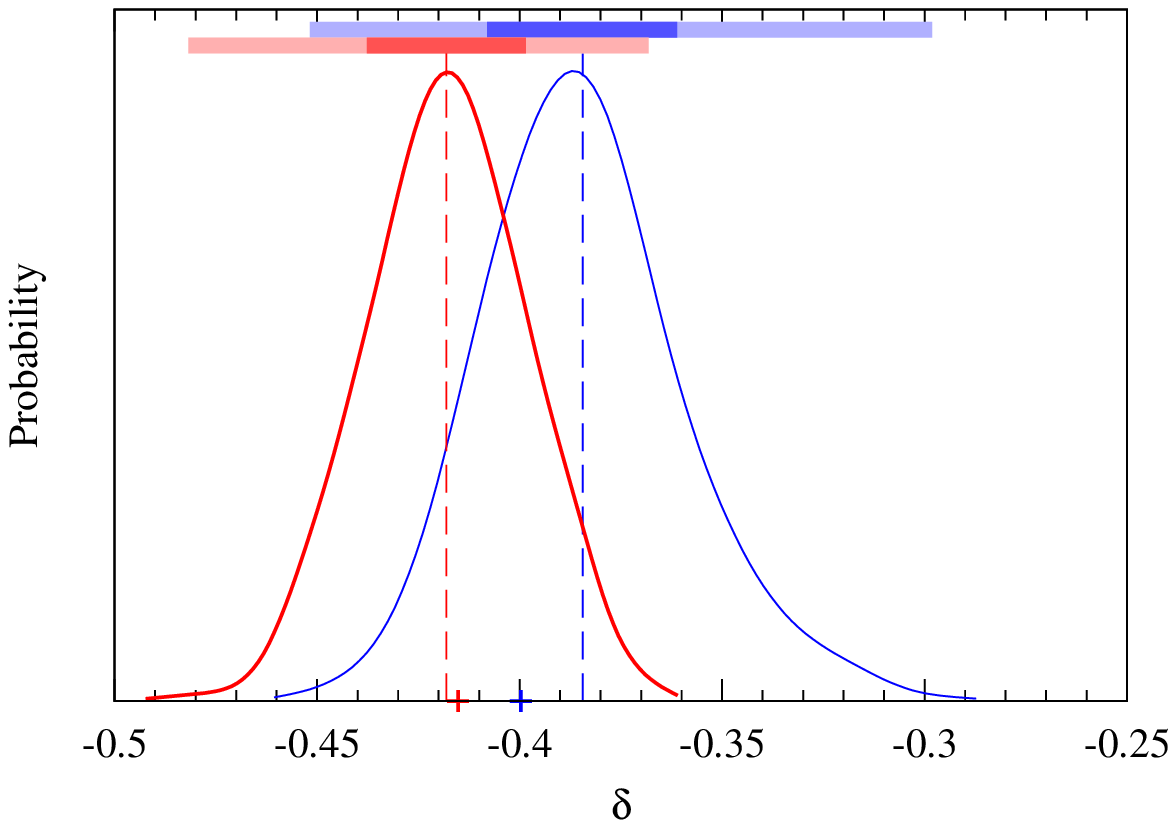}
  \caption{Top panel: $E^{3}{\rm Flux}$ as a function of energy of the electrons/positrons. The $\Phi_{\rm e^{+}}$ and $\Phi_{\rm e^{-}}$ data are taken from \cite{AMS-2014a,AMS-2014b}. Middle panel: the spectral index of $\Delta\Phi$ evolving with the energy of the electrons. Bottom panel: The probability distribution of $\delta$ found in numerical simulations with our own code \cite{Feng2014} based on the COSMOMC (http://cosmologist.info/cosmomc/). The horizontal bar indicates the $1\sigma$ and $3\sigma$ standard deviations, and the vertical dashed line (cross) represents the statistic-mean (best-fit) value. The color blue (red) represents the result of DR (DC) propagation model.}
  \label{fig:1}
\end{figure}

For such a purpose we focus on the data of $\Delta \Phi=\Phi_{\rm e^{-}}-\Phi_{\rm e^{+}}$ (see the top panel of Fig.\ref{fig:1}) that is dominated by the propagated primary CR electrons and can {\it ``minimize" the possible uncertainties of the identified excess caused by the introduction of the ``new" source(s) for the  positron excess}. Such a treatment is only possible currently thanks to the release of the AMS-02 electron/positron spectra with unprecedented accuracy in a wide energy range \cite{AMS-2014a,AMS-2014b}. The spectral index of $\Delta\Phi$ evolving with the energy of electrons is shown in the Upper right panel of Fig.\ref{fig:1} (we slide the energy window covering the energy range of every 5 neighboring data bins, within which the power law spectral index and its error are obtained) and there is not any evidence for spectral softening at $\epsilon_{\rm e}>20$ GeV where the solar modulation of cosmic ray fluxes is negligible.
It is in agreement with the empirical fit of the latest AMS-02 electron/positron data with the ``minimal model" of \cite{AMS-2013}, in which the so-called ``diffuse" electron component dominating $\Delta \Phi$ can be well approximated by a signal power-law up to the energy of $\sim 500$ GeV \cite{AMS-2014a,AMS2014-Ting}. Such a simple behavior, however, is actually {unexpected in the standard/conventional CR propagation model,} in which CRs are thought to originate in homogenously-distributed supernova remnants and the primary electrons from different sources are assumed to take a single power-law energy distribution for $\epsilon_{\rm e}>$ quite a few GeV \cite{Strong07Rev,galprop}. The higher the $\epsilon_{\rm e}$, the quicker the cooling of the diffusing electrons. The cooling timescale of electrons/positrons is $\tau_{\rm c}\sim 17~{\rm Myr}~(\epsilon_{\rm e}/10~{\rm GeV})^{-1}$ while the proton CR age is estimated to be $\tau_{\rm a}\sim 20~{\rm Myr}~(\epsilon_{\rm e}/2.6~{\rm GeV})^{-0.53}$ for $\epsilon_{\rm e}\geq 2.6~{\rm GeV}$ \cite{Strong07Rev,Piran2009}. It is reasonable to assume that the primary CR electrons and protons were from the same sources and thus at the same ages, we can then define a ``cooling" energy ($\epsilon_{\rm e,c}\sim 30~{\rm GeV}$ given by $\tau_{\rm c}=\tau_{\rm a}$) of the electrons above which the cooling softens the spectrum effectively. As a result of the superposition of the particles from different sites, the spectrum of propagated primary electrons would be continually softened.
Indeed a general behavior found in the numerical calculations is that at $\epsilon_{\rm e}>$ 10s GeV the spectrum of the propagated primary CR electrons gets softer and softer and the softening between the energy ranges of $100-400$ GeV and $10-50$ GeV is $\sim \epsilon_{\rm e}^{-0.2}$ (see for example the ``background" component of Fig.1 and Fig.2 of \cite{Feng2013}).  The inconsistence between the data and the prediction of the conventional CR model likely suggests a significant spectral excess at high energies, which could arise from for example a group of nearby supernova remnants \cite{Kobayashi2004,Vladimirov2012,Feng2014,Di2014,Gaggero2013,Boudaud2015}.

Please bear in mind that the puzzling non-softening spectral behavior of propagated primary electrons could be just an illusion if in deriving $\Delta \Phi$ either ``(a) too much electron flux has been subtracted at lower energies" or ``(b) too little electrons have been removed at high energies". If scenario (a) is correct (i.e., $\Phi_{\rm e^{+}}$ overestimates the corresponding electron flux at low energies significantly and the `intrinsic' $\Delta \Phi$ is as large as the standard CR model prediction), we need $\Phi_{\rm e^{+}}\sim 0.4\Phi_{\rm e^{-}}$ at $\epsilon_{\rm e}\sim 10$ GeV, which has already been convincingly ruled out by the ${\cal R}$ data of AMS-02. As for scenario (b), we have assumed that the sources giving rise to the positron excess component do not generate more abundant electrons at given energies, which is the case for the most widely discussed new CR-electron/positron sources include pulsars \cite{zhangli} and dark matter annihilation/decay \cite{Arkani-Hamed2009,Decay2009}, for which the electrons/positrons were born in pairs (One exception is the so-called asymmetric dark matter model, in which the possibility of decaying into electrons and positrons does not equal with each other \cite{FengKang2013}). Moreover, for the collision of high energy CRs with other particles/photons taking place in both the interstellar medium and the CR sources, it is well known that among the resulting secondary particles the positrons are more (rather than less) than electrons \cite{Strong07Rev,fan,Blasi2009}. For instance, the most-widely discussed proton$-$proton and proton$-$Helium collisions in the interstellar medium (these processes have also been properly taken into account in our numerical fit of $\Delta \Phi$, see below) yield charged pions and kaons, which further decay as $K^{\pm}\rightarrow \pi^{\pm}+\pi^{0},~K^{\pm}\rightarrow \mu^{\pm}+\nu_{\mu},~\pi^{\pm}\rightarrow \mu^{\pm}+\nu_{\mu}$ and $\mu^{\pm}\rightarrow e^{\pm}+\bar{\nu}_{\mu}+\nu_{\rm e}$. At $\epsilon_{\rm e}\gg 1$ GeV, the secondary electrons have a flux about half of the corresponding positrons \cite{Strong07Rev,fan}. Hence the hypothesis described in scenario (b) does not apply, either. So far we have shown that the non-softening spectral behavior of propagated primary electrons is reliable.

\begin{table}[!htb]
\caption {The propagation parameters.}
\begin{tabular}{ccccc}
\hline \hline
&& DR &&DC\\
\hline
$z_{h}{\rm (kpc)}$ &&4&&{2.5}\\
$D_{0}$($10^{28}$ cm$^2$ s$^{-1}$) &&5.30&&{ 1.95}\\
diffusion index\footnotemark[1]($\delta^\prime_1/\delta^\prime_2$) &&0.33/0.33&&{ 0/0.51}\\
$v_A$(km s$^{-1}$) &&33.5&&/\\
$dV_{c}/dz$(km s$^{-1}$ kpc$^{-1}$) &&/&&{4.2}\\
$p$ injection\footnotemark[2]($\gamma_1/\gamma_2$) &&1.88/2.39&&{1.88/2.39}\\
$E_{\rm br}$(GeV) &&11.5&&{7.4}\\
\hline
\hline
\end{tabular}
\label{tablee}
\\\footnotemark[1]{Below/above rigidity $\rho_0={4.71}$ GV.} \\
\footnotemark[2]{Below/above $E_{\rm br}$.}
\end{table}

The propagation of CR can be described by a {transport equation} including diffusion, convention, re-acceleration, radioactive and so on \cite{Strong07Rev}.
As usual we adopt the GALPROP \cite{galprop} package to calculate the propagation of the CR particles numerically.
The diffusion-reacceleration (DR) and diffusion-convection (DC) model are introduced to discuss the systematic uncertainty of CR propagation. The CR propagation parameters are fixed in our discussion, which can reasonably fit the observational ${\rm B/C}$, ${\rm ^{10}B{\rm e}/^{9}B{\rm e}}$ and proton data.
To be precise, we use parameters in \cite{ptuskin2006,ackermann2012} when discussing DR model, while we fix the propagation parameters \cite{LinSJ2015,liujie} and fit the latest AMS-02 proton data \cite{AMS-2015} to get proton injection parameters in DC model.
The main parameters we used are summarized in Table. \ref{tablee}. To account for the possible spectrum ``hardening" of the injected primary electrons, three spectral indexes ($\Gamma_1, \Gamma_2, \Gamma_3$) and two break rigidities ($\rho_{\rm br1}, \rho_{\rm br2}$) are assumed in the numerical modeling. Note that the first break rigidity is introduced to interpret the data that is about $10~{\rm GV}$. Though we constrain all the two break parameters but here we just discuss the origin of the second one. A parameter $\delta\equiv \Gamma_3 - \Gamma_2$ is defined to describe the possible spectral change. The case of $\delta < 0~(>0)$ refers to the energy spectral hardening (softening) at the break rigidity $\rho_{\rm br2}$.

\begin{figure}
\includegraphics[width=80mm,angle=0]{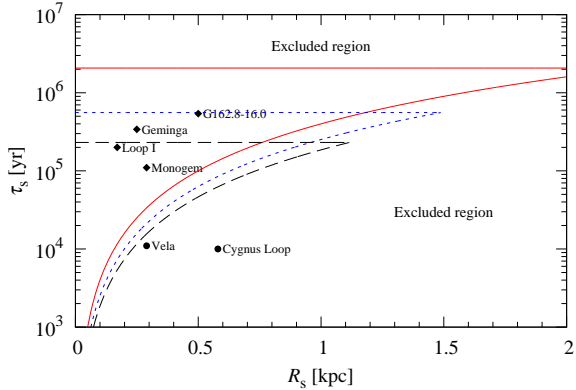}
  \caption{Some nearby possible sources for the primary electron excess. The regions covered by solid, dotted and dashed lines are for the sources of electrons at energies of 100 GeV, 400 GeV and 1 TeV, respectively. We assume that no energetic electrons are effectively accelerated if the sources are older than $\sim 10^{5}$ yr.}
  \label{fig:2}
\end{figure}

In this work we use Markov Chain Monte Carlo (MCMC) methods to determine the probability distribution function (PDF) of the posterior model parameters by sampling the distribution according to the prior PDF and the
likelihood function. The code was developed by ourselves in \cite{Feng2014}.
The MCMC sampler we used here is COSMOMC basing on Metropolis-Hastings algorithm. The data used to calculate the likelihood is shown in the upper panel of Figure \ref{fig:1}.  The free parameters to be fitted are $\{\Gamma_1, \Gamma_2, \delta, \rho_{\rm br1}, \rho_{\rm br2}, N_{\rm e}, \phi\}$, where $N_{\rm e}$ is the normalized electron flux at 25 GeV, and $\phi$ is the potential of solar modulation. $c_{\rm e^+}$, the factor used to re-scale the absolute fluxes of secondary particle electrons and positrons due to the uncertainties in the calculation of CR secondary particles, is set to be $1$.
The fit parameters of the AMS-02 $\Delta \Phi$ data is shown in Table \ref{table1} for DR model and in Table \ref{table2} for DC model. The best-fit yields $\delta=-0.40$ ($-0.42$) and $\rho_{\rm br2}=50.1~ {\rm GeV}$ ($49.1~ {\rm GeV}$) for DR (DC) model, the corresponding minimal $\chi^2/{\rm d.o.f}$ is $0.81$(0.52) and we call it our global best fit model, where ${\rm d.o.f}$ represents the degree of freedom. From the calculation, we can see that the best fit of $\delta$ and $\rho_{br2}$ are consistent with each other in the two CR propagation models. But if we set $\delta =0$, the minimal $\chi^2/{\rm d.o.f}$ we got is $2.0$  ($2.18$), too large to be acceptable.
The 1-D marginalized posterior PDFs of $\delta$ is shown in bottom panel of Fig.\ref{fig:1}. In particular at the confidence level of $99.7\%$ we have $\delta= -0.38^{+0.09}_{-0.07}$($-0.42^{+0.05}_{-0.06}$) and $\rho_{br2}=52.4^{+14.5}_{-8.3}$($49.5^{+12.6}_{-8.5}$) in DR (DC) model. All $\delta$ are smaller than $-0.25$ after burn-in $50\%$ of the samples, implying that the case of $\delta=0$ (i.e., without ``hardening" or equally ``primary electron excess") has been convincingly ruled out (the confidence level is well above $5\sigma$).
Our global best fit to $\Delta \Phi$ is shown in the top panel of Fig.1. The amplitude of the primary electron excess component can be estimated by subtracting the ``background" component from $\Delta\Phi$, where the ``background" represents the theoretical flux of $\Delta \Phi$ predicted in our global best fit model except setting $\delta=0$ (see Fig.1: the case 1).
At energies of $\sim 100-400$ GeV, the primary electron excess component has a flux that is about twice of the positron excess component\footnote{We refit the cosmic ray electron and positron data with an additional "symmetric" source component together with the background component. We then remove the background component from the $e^{+}$ data to get the excess component.}
(see Fig.1: the case 1). If the primary electron excess component is absent, the increasing of ${\cal R}$ at energies $>100$ GeV would be (much) quicker than that measured by AMS-02 and ${\cal R}$ will peak at $\sim 30\%$ (i.e., about twice of the observed peak value).

\begin{table}[!htb]
\begin{small}
\caption {The fit parameters of the AMS-02 $\Delta\Phi$ data for DR model.}
\begin{tabular}{ccc}
\hline \hline
 & Including Hardening & No Hardening  \\
 & Best Fit/Posterior mean/$3\sigma$ range& Best Fit\\
\hline
$\Gamma_1$ & 2.188/2.143/[1.976,2.266] & 1.490 \\
$\rho_{\rm br1}({\rm GV})$ & 6.183/6.041/[5.254,7.016] & 5.536  \\
$\Gamma_2$ & 3.059/3.032/[2.933,3.115] & 2.727 \\
 $\rho_{\rm br2}({\rm GV})$  &50.073/52.449/[44.169,66.905]  & -  \\
$\delta$ & -0.400/-0.384/[-0.452,-0.298] & -\\
$N_{\rm e}^1$ & 1.432/1.416/[1.335,1.493]  & 1.191  \\
$\phi({\rm GV})$ & 1.452/1.387/[1.124,1.626] &  0.579 \\
$\chi^2/{\rm d.o.f}$&0.81 & 2.0\\
  \hline
  \hline
\end{tabular}
\label{table1}\\
\end{small}
  $^1$ In this work $N_e$ is in unit of $10^{-9}~{\rm cm^{-2}~sr^{-1}~s^{-1}~MeV^{-1}}$.
\end{table}

\begin{table}[!htb]
\begin{small}
\caption {The fit parameters of the AMS-02 $\Delta\Phi$ data for DC model.}
\begin{tabular}{ccc}
\hline \hline
 & Including Hardening & No Hardening  \\
 & Best Fit/Posterior mean/$3\sigma$ range& Best Fit\\
\hline
$\Gamma_1$ & 1.061/1.624/[0.802,3.396] &  1.636 \\
$\rho_{\rm br1}({\rm GV})$ & 2.016/2.003/[1.010,6.943] &  5.519  \\
$\Gamma_2$ & 3.0~~/2.999/[2.948,3.057] &  2.673 \\
 $\rho_{\rm br2}({\rm GV})$  &49.113/49.523/[40.994,62.097]  & -  \\
$\delta$ & -0.415/-0.418/[-0.482,-0.368] & -\\
$N_{\rm e}^1$ & 1.391/1.390/[1.368,1.415]  &  1.198  \\
$\phi({\rm GV})$ & 1.129/1.127/[1.053,1.208] &  0.378  \\
$\chi^2/{\rm d.o.f}$&0.52 & 2.18 \\
  \hline
  \hline
\end{tabular}
\label{table2}\\
\end{small}
\end{table}

The introduction of a global spectral hardening with $\delta \approx -0.4$ for all injected primary electrons yields much more high energy particles than the case of $\delta=0$ and the cooled ones ``pile up" at lower energies. Consequently, the low energy spectrum gets hardened indirectly, which in turn renders the injection spectrum softened in the modeling. Hence the flux of the primary electron excess obtained above (i.e., case 1) might have been overestimated. Such a fact motivates us to have a more ``conservative" estimate on the excess component. Since the whole set of $\Delta \Phi$ data can not be reasonably fitted within the standard/convential CR propagation model, in the new approach only the data at energies of $\epsilon_{\rm f}\leq 50$ GeV (slightly below $\rho_{\rm br2}$) are modeled. The underlying assumption is that the cooling of the excess component is not efficient enough to play a substantial role in ``hardening" the low energy electron spectrum, which is the case if the excess component is dominated by some nearby and relatively-young sources. The standard CR propagation model (for simplicity, here we just consider the DR model) can well reproduce such an ``incomplete" set of data and the best fit gives $\chi^{2}/{\rm d.o.f}=1.03$ (the best fit parameters are $\{\Gamma_1, \Gamma_2, \rho_{\rm br1}, N_{\rm e}, \phi\}\sim \{1.990, 2.922, 5.577~{\rm GV}, 1.364,~1.169~{\rm GV}\}$, respectively). The extrapolation of the ``background" flux to energies $>\epsilon_{\rm f}$ is significantly below the data, suggesting a distinct excess with a flux somewhat smaller than that in case 1, as expected (see the top panel of Fig.1: case 2).

We conclude that with respect to the prediction of the ``standard/conventional CR propagation model" there is a distinct primary CR electron excess in the AMS-02 $\Delta \Phi$ data and its  flux is comparable to that of the positron excess at energies of $\sim 100-400$ GeV. It has to be properly taken into account in the modeling of the CR electron/positron data within the standard/convential CR propagation scenario, otherwise the inferred physical parameters of the new positron sources (e.g., dark matter particles or pulsars) would be biased \cite{Feng2014,pulsar2013}. The physical origin of such a new excess component, however, is hard to pin down uniquely. Among  various possibilities, we think that some nearby middle-aged sources in particular supernova remnants \cite{Atoyan1995,Kobayashi2004,Piran2009,Vladimirov2012,Di2014,Gaggero2013,Boudaud2015} may play the leading role. The requests of both ``nearby" and ``middle-aged" are for the following reasons: (i) The electrons at energies of trans-TeV and beyond lost their energy very quickly and hence can reach us only if the sources are at a radius  $R_{\rm s} \leq 1~{\rm kpc}~({\epsilon_{\rm e}/1~{\rm TeV}})^{-1/3}$ \cite{Strong07Rev,fan}; (ii) The presence of distinct primary electron excess at $\sim 100$ GeV requires that such particles have transported to us, requiring a lifetime of the sources $\tau_{\rm s} \geq \tau_{\rm d}\equiv 4\times 10^{5}~{\rm yrs}~(R_{\rm s}/1~{\rm kpc})^{2}(\epsilon_{\rm e}/100~{\rm GeV})^{-1/3}$. Too old sources however are disfavored due to the dilution of the flux of the CRs as a function of time ($\propto \tau_{\rm s}^{-3/2}$) and due to the quick cooling of the electrons. Geminga with $(\tau_{\rm s},~R_{\rm s})\sim (3.4\times 10^{5}~{\rm yr},~0.25~{\rm kpc})$ \cite{Kargaltsev2008}, Monogem with $(\tau_{\rm s},~R_{\rm s})\sim (1.1\times 10^{5}~{\rm yr},~0.29~{\rm kpc})$ \cite{Shibanov2006}, Loop I with $(\tau_{\rm s},~R_{\rm s})\sim (2\times 10^{5}~{\rm yr},~0.17~{\rm kpc})$ \cite{Egger1995}  and G 162.8$-$16.0 with $(\tau_{\rm s},~R_{\rm s})\sim (5.4\times 10^{5}~{\rm yr},~0.5~{\rm kpc})$ \cite{DeLuca2011} are suitable candidates of discrete instantaneous sources for the primary electron excess (see Fig.\ref{fig:2}, in which the cooling rates of electrons at different energies are taken from \cite{Kobayashi2004}; see also \cite{Piran2009} for illustrative calculation). In particular, Monogem may be the dominant source for the identified excess that might hold to $\sim 1$ TeV. While some nearby but `young' (i.e., $\tau_{\rm s}<\tau_{\rm d}$) supernova remnants such as Cygnus Loop with $(\tau_{\rm s},~R_{\rm s})\sim (10^{4}~{\rm yr},~0.58~{\rm kpc})$ \cite{Blair2009} and Vela with $(\tau_{\rm s},~R_{\rm s})\sim (1.1\times 10^{4}~{\rm yr},~0.29~{\rm kpc})$ \cite{Kargaltsev2008} may give rise to TeV-PeV excess possibly in both electron spectrum and nuclei spectra since only such high energy particles might have reached us \cite{Kobayashi2004}. Due to its quite uncertain $R_{\rm s}$ and $\tau_{\rm s}$ \cite{Shinn2006}, the role of Lupus Loop is less clear. Other physical processes that could (partly) account for the primary electron excess include the injection spectrum hardening at high energies (as expected in the non-linear CR acceleration model \cite{Ptuskin2013}) and the superposition of the variable injection spectra of the CR sources (i.e., some sources can accelerate CRs with harder spectra than the typical \cite{lab4}). In the model of nearby discrete supernova remnants, multiple sub-structures in the excess spectrum and some anisotropy of the 100s GeV electrons are expected. In the models of both non-linear CR acceleration and superposition of the variable injection spectra of the CR sources, similar excesses seem ``unavoidable" in the nuclei spectra.
Hence, the self-consistent modeling of the upcoming CR nuclei data by AMS-02 and other space missions may shed valuable light on the physical origin of the primary electron excess identified in this work and  `localize'/identify some nearby cosmic ray sources.

Finally, we would like to point out that the hardening of the electron spectrum could also be caused by an abrupt ``decrease of the diffusion index" \cite{Gaggero2013}. If correct, similar spectral hardening would appear also in proton, helium and B/C data. So far the helium and B/C data have not been officially published by the AMS-02 collaboration, yet. The proton data indeed shows a spectral hardening at $\sim 340$ GV \cite{AMS-2015}, which however seems to be (sizably) higher than the electron break ($\sim 50$ GV) inferred in this work. Nevertheless, we plan to examine whether the ``diffusion-index change" model can interpret the electron/positron data, the proton/helium data, and the B/C data self-consistently when all these data have been officially published by the AMS-02 collaboration.

{\it Acknowledgments:}  This work was supported in part by 973 Program of China under grant 2013CB837000,  National Natural Science of China under grants 11303096, 11173064 and 11233001, and the Foundation for
Distinguished Young Scholars of Jiangsu Province, China (No. BK2012047). This work is in memory of Prof. Tan Lu, one of the founders of high energy astrophysics researches in China, who passed away on 2014 December 3.
\\


\end{document}